# LMI Approach for Sliding Mode Control and Analysis of DC-DC Converters

ALEKSANDRA D. LEKI  , University of Belgrade,
  School of Electrical Engineering, Belgrade
DUŠAN M. STIPANOVI  , University of Illinois,
  Coordinated Science Laboratory, Urbana, IL USA

*Circuits' and in particular DC/DC converters' switching behavior is analyzed in this paper using the equivalent control modeling of the dynamic systems' sliding mode regime. As a representative example and also being one of the most complex circuits among DC/DC converters, the   uk converter is chosen. It is shown how the converter's behavior in the steady state regime can be studied and analyzed by the linear matrix inequalities based stability conditions for linear dynamic systems with nonlinear sector bounded perturbations. The maximization of the nonlinear sector bound provides a limit for applying the linear ripple approximation in the converter operation analysis. Furthermore, our approach is validated by providing simulation results for two different switching surfaces of practical interest.*

**Key words**: *DC/DC converter,   uk converter, sliding mode, linear matrix inequalities*

## 1. INTRODUCTION

In this work, we study a sliding mode regime as a nonlinear (due to switching) system modeling behavior [1-2] of DC/DC converters using the concept of equivalent control as introduced and developed by Utkin (see, for example, [1, 3]). In particular, we consider the   uk converter as a representative, one of the most complex, and important example of DC/DC converters [12]. We propose an approach in which a nonlinear dynamic model resulting from applying so called equivalent control [1], is linearized around its equilibria and represented as the sum of its linearized portion and the remainder which is nonlinear. This representation is known to be well suited for a Lyapunov stability analysis based on Linear Matrix Inequalities (LMI) [4].

LMI represent a special class of convex optimization formulations [5] which thus inherit all the benefits of convex programming algorithms such as uniqueness of solutions, convergence to a solution if it exists, and a definite test for feasibility of problems where the answer depends on whether or not convex algorithms converge. In particular, the LMI formulation enables an application of convex programming in computing a quadratic Lyapunov function which proves stability of the overall nonlinear system as well as switched systems [6].

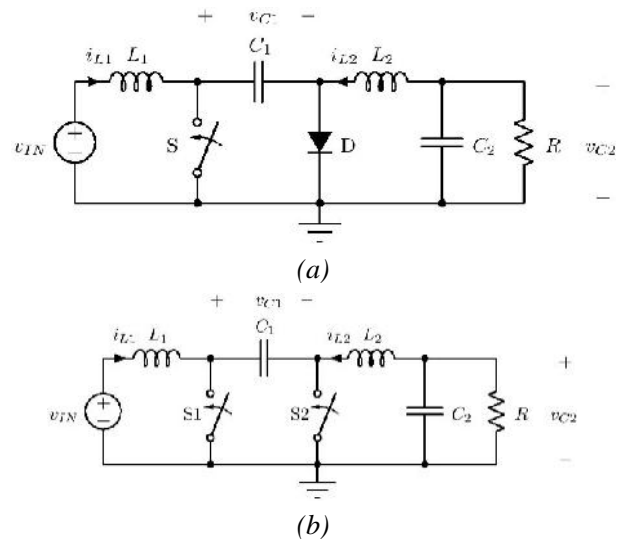

*Figure 1 –   uk converter containing (a) unidirectional switch S and a diode D; (b) bidirectional switches S1 and S2.*

One of the key ingredients in this formulation is the sector bounding of the additive nonlinearity which is thus treated as a perturbation [7]. Furthermore we obtain a relationship between the simulation results and the domain in which the nonlinearity is sector bounded with the sector being maximized using LMI formulation. This relationship enables us to estimate a size of the hysteresis used to design a switching sequence for which the converter operates in a proper steady state regime.

Most of the research on control of the  uk converter has been focusing on controlling two state variables, as in [8] where it is related to the sliding mode control under an assumption that $C_1 \gg C_2$ (for the capacitors $C_1$ and $C_2$ denoted as in Fig. 1) which results in an overall system reduction.

This reduction was performed in some particular cases by exactly one reactive element elimination [9]. However, there are some other works concerning sliding mode control of the  uk converter which incorporated a full-order system model as reported in [10].

Publication [10] describes sliding mode controls of the  uk converter built with bidirectional switches, which do not produce any discontinuous regimes. LMI approach has been already used for a design of the sliding-mode controlled buck converter [11].

In this paper we show how the sliding mode control of the  uk converter in its all continuous and discontinuous regimes can be analyzed using an LMI based stability analysis for linear systems with an additive nonlinearity. Stability conditions in the case of each switching surface are formulated using LMI and the additive nonlinearity (treated as a perturbation) bound is computed. The sector bounded nonlinearity is shown to be in a direct connection with the linear ripple approximation [12].

The paper is organized as follows. In Section 2, the sliding mode analysis and how to compute equivalent control for a class of dynamic systems which model DC/DC circuits' switching behavior are presented. Dynamic properties of the  uk converter and its modeling are described and developed in Section 3. In Section 4, a switching surface analysis for two different surfaces of interest as well as corresponding simulation results are provided. Finally, some concluding remarks are summarized in Section 5.

## 2. SLIDING MODE ANALYSIS, STABILITY AND CONTROL

For a switching system such as DC/DC converter [13], the concept of the sliding mode control is materialized through an ON-OFF operation of a controllable switch. Considering that for any converter there is only one switching signal, there are two subsystems in the continuous conduction mode which can be rewritten in a compact matrix form as

$$\dot{\mathbf{x}} = \mathbf{A}\mathbf{x} + \mathbf{B} + (\mathbf{C}\mathbf{x} + \mathbf{D})u \quad (1)$$

where $\mathbf{x}$ denotes the state vector consisting of $n$ state variables and $u \in \{0,1\}$ is the scalar indicator showing whether the controllable switch is conducting or not. Matrices A, B, C and D are of appropriate dimensions determined by dimensions of $x$ and $u$. Essentially the system (1) represents a set of $n$ linear differential equations with a switching variable. For the control of DC/DC converters, the switching law is given by

$$u = \begin{cases} 1, & \text{switch to 1 when } S(\mathbf{x}) < -\Delta \\ 0, & \text{switch to 0 when } S(\mathbf{x}) > \Delta, \end{cases} \quad (1)$$

where we assume that the sliding surface is of a linear form $S(\mathbf{x}) = \mathbf{M}(\mathbf{x} - \mathbf{x}^*)$ with the coefficients being entries of a row vector $\mathbf{M} = [m_1 \cdots m_n]$, $\Delta$ is a positive number and a constant column vector $\mathbf{x}^*$ is an equilibrium point and steady-state vector.

Sliding mode control technique is applied in order to analyze the converter's behavior on the switching surface using the concept of equivalent control [1].

The system's motion in the sliding mode is restricted to the switching surface $S(\mathbf{x}) = 0$ if the surface is attractive. In the steady-state is also $\dot{S}(\mathbf{x}) = 0$, [1], and thus, the equivalent control can be obtained from the vector equation

$$u_{eq} = -\left(\frac{\partial S}{\partial \mathbf{x}}(\mathbf{C}\mathbf{x} + \mathbf{D})\right)^{-1} \frac{\partial S}{\partial \mathbf{x}}(\mathbf{A}\mathbf{x} + \mathbf{B}), \quad (2)$$

if $\frac{\partial S}{\partial \mathbf{x}}(\mathbf{C}\mathbf{x} + \mathbf{D})$ is nonsingular, which will be satisfied in the cases of our interest. After substituting $u_{eq}$ in Eq. (1) we obtain

$$\dot{\mathbf{x}} = \mathbf{A}\mathbf{x} + \mathbf{B} + (\mathbf{C}\mathbf{x} + \mathbf{D})u_{eq}. \quad (4)$$

An equilibrium and steady-state vector $\mathbf{x}^*$ can now be computed as a constant solution to Eq. (4). Furthermore, we represent the system as the sum of the linear part $\mathbf{A}_1(\mathbf{x} - \mathbf{x}^*)$ (computed using Taylor's expansion) and the exact nonlinear remainder computed as

$$\mathbf{h}(\mathbf{x}) = \dot{\mathbf{x}} - \mathbf{A}_1(\mathbf{x} - \mathbf{x}^*) \quad (5)$$

where $\dot{\mathbf{x}}$ is given in the Eq. (4).

The sliding surface equation $S(\mathbf{x}) = 0$ enables an elimination of one state variable, in general. By introducing the change of variables $\mathbf{y} = \mathbf{x} - \mathbf{x}^*$, where $\mathbf{x}^*$ is an equilibrium, and denoting as $\mathbf{z}$ the vector of variables $y_i$ which are not eliminated, the reduced system dynamics can be written as

$$\dot{\mathbf{z}} = \mathbf{A}^*_{(n-1)\times(n-1)} \mathbf{z} + \mathbf{h}^*_{(n-1)\times 1}(\mathbf{z}), \quad (6)$$

where matrix $\mathbf{A}^*$ and nonlinearity $\mathbf{h}^*$ are obtained after one state variable is eliminated using the sliding surface equation.

System stability on the desired sliding surface is determined by finding a Lyapunov function candidate of the quadratic type

$$V(\mathbf{z}) = \mathbf{z}^T \mathbf{P} \mathbf{z}, \quad (3)$$

where $\mathbf{P}$ is a constant positive definite matrix. The next step is to compute $\dot{V}(\mathbf{z})$ as follows:

$$\dot{V}(\mathbf{z}) = \mathbf{z}^T (\mathbf{A}^{*T}\mathbf{P} + \mathbf{P}\mathbf{A}^{*T}) \mathbf{z} + 2\mathbf{z}^T \mathbf{P} \mathbf{h}^*(\mathbf{z}) \quad (4)$$

The stability analysis is performed using LMI [4] which is an efficient and almost the only way to compute Lyapunov functions for systems whose dynamics are represented by a linear and a sector bounded nonlinear part. The LMI problem is formulated using the S-procedure [14] which maximizes the bound on the nonlinear part. This approach due to being based on convex programming will always provide a solution if the problem is feasible and the convergence in the case when the solution exists is fast. Now, the LMI problem to examine systems stability can be formulated as follows [7]:

$$\begin{array}{c}\text{minimize } x \\ \text{subject to } \mathbf{Y} > 0 \\ \begin{bmatrix} \mathbf{A}^*\mathbf{Y} + \mathbf{Y}\mathbf{A}^{*T} & \mathbf{I} & \mathbf{Y}\mathbf{H}^T \\ \mathbf{I} & -\mathbf{I} & \mathbf{0} \\ \mathbf{H}\mathbf{Y} & \mathbf{0} & -x\mathbf{I} \end{bmatrix} < 0. \end{array} \quad (5)$$

A solution would provide matrix $\mathbf{Y} = \ddagger \mathbf{P}^{-1}$, for some positive value (see [7] for more details), which maximizes the sector bound $r = \dfrac{1}{\sqrt{x}}$ so the system is Lyapunov stable. Lyapunov function derivative is then

$$\dot{V}(\mathbf{z}) = -\mathbf{z}^T \mathbf{Q} \mathbf{z} + 2\mathbf{z}^T \mathbf{P} \mathbf{h}^*(\mathbf{z}) < 0, \mathbf{z} \neq 0, \quad (6)$$

where $\mathbf{Q} = -(\mathbf{A}^{*T}\mathbf{P} + \mathbf{P}\mathbf{A}^{*T}) < 0$ and the nonlinear term/remainder satisfies the following sector bound

$$\mathbf{h}^{*T}\mathbf{h}^* \leq r^2 \mathbf{z}^T \mathbf{H}^T \mathbf{H} \mathbf{z}. \quad (7)$$

The nonlinear remainder's sector bound shape is determined by a matrix $\mathbf{H}$ and a positive parameter $r > 0$. The following nonsingular transformation [7]

$$\mathbf{x} = \mathbf{T}\tilde{\mathbf{x}}, \quad (8)$$

with matrix $\mathbf{T}$ being formed from the eigenvectors of the matrix $\mathbf{A}^*$ provides larger values for , that is, a better sector bound. Therefore, the LMI formulation in the transformed space

$$\tilde{\mathbf{H}} = \mathbf{H}\mathbf{T}, \quad (9)$$

for the preselected matrix $\tilde{\mathbf{H}}$ results in sector bound value denoted as $\tilde{r}$, which is related to as

$$r = \dfrac{\tilde{r}}{\|\mathbf{T}^{-1}\|}. \quad (10)$$

The maximization of the sector bound provides a limit for applying the linear ripple approximation, that is, a limit for considering state variables time constants to be negligible.

## 3. MODELING BEHAVIORS OF THE ĆUK CONVERTER

The Ćuk converter depicted in Fig. 1 is a complex fourth order DC/DC converter [12] which for those reasons is chosen to illustrate our approach. In particular, Figure 1a shows a Ćuk converter which applies switches realized by an unidirectional switch and a diode, while Figure 1b shows a synchronous Ćuk converter employing current and voltage bidirectional switches. For the synchronous converter bidirectional switches are controlled with state(S2)=¬state(S1), where ¬ denotes negation.

### A. Sliding domain and operating modes

Depending on the Ćuk converter's switches realization, a few operating modes can be determined and analyzed. Both realizations of the Ćuk converter depicted in Fig. 1 operate in two continuous modes, that is, the first one being when switch S is on and a diode isn't (in Figure 1b switch S1 is on and S2 is off) and the second one being when switch S is off and diode D is conducting (switch S1 is off and S2 is on where again we refer to Figure 1b).

In the continuous conduction mode the state space equations are given by

$$\dfrac{di_{L1}}{dt} = \dfrac{v_{IN}}{L_1} - \dfrac{v_{C1}}{L_1}(1-u), \quad \dfrac{di_{L2}}{dt} = \dfrac{v_{C1}}{L_2}u + \dfrac{v_{C2}}{L_2},$$
$$\dfrac{dv_{C1}}{dt} = \dfrac{i_{L1}}{C_1}(1-u) - \dfrac{i_{L2}}{C_1}u, \quad \dfrac{dv_{C2}}{dt} = -\dfrac{i_{L2}}{C_2} - \dfrac{v_{C2}}{RC_2}, \quad (15)$$

where $u = 1$ is when S is on and D is off and $u = 0$ when S is off and diode D is on. Additionally, switch S1 from Fig. 1b is controlled with signal $u$ and switch

S2 with $\bar{u}$. Eq. (15) can be rewritten in the compact matrix form suitable for applying sliding mode control as in Eq. (1), with matrices defined as follows:

$$\mathbf{A} = \begin{bmatrix} 0 & 0 & -\frac{1}{L_1} & 0 \\ 0 & 0 & 0 & \frac{1}{L_2} \\ \frac{1}{C_1} & 0 & 0 & 0 \\ 0 & -\frac{1}{C_2} & 0 & -\frac{1}{RC_2} \end{bmatrix}, \mathbf{B} = \begin{bmatrix} \frac{v_{IN}}{L_1} \\ \frac{v_{IN}}{L_2} \\ 0 \\ 0 \end{bmatrix}, \mathbf{C} = \begin{bmatrix} 0 & 0 & \frac{1}{L_1} & 0 \\ 0 & 0 & \frac{1}{L_2} & 0 \\ -\frac{1}{C_1} & -\frac{1}{C_1} & 0 & 0 \\ 0 & 0 & 0 & 0 \end{bmatrix}$$
(11)

with matrix **D** being zero matrix and thus omitted in what follows. Vector of the state space variables $\mathbf{x} = [i_{L1}\ i_{L2}\ v_{C1}\ v_{C2}]^T$ is defined as $x_1 = i_{L1}$, $x_2 = i_{L2}$, $x_3 = v_{C1}$ and $x_4 = v_{C2}$, with the prescribed syntax steady-state values being $\overline{i_{L1}} = x_1^*$, $\overline{i_{L2}} = x_2^*$, $\overline{v_{C1}} = x_3^*$ and $\overline{v_{C2}} = x_4^*$.

A synchronous converter depicted in Figure 1b can operate only in the continuous modes. Additionally, the uk converter shown in Figure 1a operates in two discontinuous modes, discontinuous inductor current mode (DICM) and discontinuous capacitor voltage mode (DCVM).

DICM occurs when the switch S is turned off and the sum of the inductor currents goes to zero, that is, $i_{L1} + i_{L2} = 0$, which turns off the diode. State space equations in DICM are as follows:

$$\frac{di_{L1}}{dt} = -\frac{di_{L2}}{dt} = \frac{v_{IN}}{L_1 + L_2} - \frac{v_{C1}}{L_1 + L_2} - \frac{v_{C2}}{L_1 + L_2},$$
$$\frac{dv_{C1}}{dt} = -\frac{i_{L2}}{C_1}, \qquad \frac{dv_{C2}}{dt} = -\frac{i_{L2}}{C_2} - \frac{v_{C2}}{RC_2}. \quad (12)$$

DCVM occurs in the state when swith S is on and the diode is off and capacitor voltage $v_{C1}$ reaches zero. That creates conditions for the diode to turn on and converter enters DCVM resulting in the following state space equations:

$$\frac{di_{L1}}{dt} = \frac{v_{IN}}{L_1},\quad \frac{di_{L2}}{dt} = \frac{v_{C2}}{L_2},$$
$$\frac{dv_{C1}}{dt} = 0,\quad \frac{dv_{C2}}{dt} = -\frac{i_{L2}}{C_2} - \frac{v_{C2}}{RC_2}. \quad (18)$$

In order to apply the sliding mode control technique to the uk converter, inverting hysteresis regulator with thresholds $\pm \Delta$ is used.

Regulator produces controlling signal $u$ as defined in Eq. (2) with the sliding surface determined using the following equation:

$$S(x) = m_1 x_1 + m_2 x_2 + m_3 x_3 + m_4 x_4 - m_5 \quad (13)$$

where $m_5$ becomes a weighted sum of the reference values multiplied by the corresponding coefficients $m_1,\cdots,m_4$ given with expression

$$m_5 = m_1 x_1^* + m_2 x_2^* + m_3 x_3^* + m_4 x_4^*. \quad (14)$$

We primarily consider a uk converter containing a switch and a diode which can operate in all continuous and discontinuous modes. For such a converter, state variable $x_3$ is positive and $x_4$ is negative, in the steady state, respectively. Thus, it is assumed that $m_1 > 0$, $m_2 > 0$, $m_3 > 0$, $m_4 < 0$ and $m_5 > 0$. Now, an equivalent control is $u_{eq} = -x_4^* / (v_{IN} - x_4^*)$ and the corresponding steady-state $x_1^* = x_4^{*2}/(Rv_{IN})$, $x_2^* = -x_4^*/R$, $x_3^* = v_{IN} - x_4^*$, where $x_4^*$ is determined by the switching surface.

*B. Ripple approximation*

The characteristic values of the DC/DC converters' are adjusted so that the time constants of the capacitors and the inductors are much longer than the switching period. Thus, the capacitors' voltages and inductors' currents steady-state values are considered when the ripple is computed [12]. This means that, by using the linear ripple approximation, inductors' currents of the uk converter either linearly increase or decrease during a switching period. When the switch S is on, for $u_{eq}T_S$ time interval, then the inductors' currents increase as

$$\Delta i_{L1} = \frac{v_{IN}}{L_1} u_{eq} T_S,$$
$$\Delta i_{L2} = \frac{x_3^* + x_4^*}{L_2} u_{eq} T_S = \frac{v_{IN}}{L_2} u_{eq} T_S, \quad (21)$$

while when the diode is on during the rest of the period, $(1-u_{eq})T_S$ long, the currents decrease by the same amount. Similarly, for the capacitors' currents we apply linear ripple approximation [12] and obtain that capacitor $C_1$'s voltage ripple is equal to

$$\Delta v_{C1} = x_2^* u_{eq} \frac{T_S}{C_1}, \quad (15)$$

while the voltage ripple on capacitor $C_2$ is very small comparing to absolute ripple change of the inductors' currents $\Delta i_{L1}$ and $\Delta i_{L2}$ and capacitor voltage $\Delta v_{C1}$. Capacitor $C_2$'s current during the switching period equals $i_{C2} = -i_{L2} - \frac{v_{C2}}{R}$, so the capacitor's voltage variation is

$$\Delta v_{C2} = \frac{\Delta i_{L2}}{8 C_2} T_S. \quad (16)$$

Using the steady state equations and the ripple of the state variables given in Eqs. (21), (22) and (23) one can determine the hysteresis value.

## 4. SWITCHING SURFACE ANALYSIS

In this section we are going to provide analysis for the Ćuk converters' behavior on the switching surfaces $S(\mathbf{x}) = x_1 - m_5$ and $S(\mathbf{x}) = m_1 x_1 + m_2 x_2 - m_5$ as representative ones, yet the procedure is general and thus can be applied to any other switching surface in the same way. One of the surfaces of the importance is $S(\mathbf{x}) = m_1 x_1 + m_4 x_4 - m_5$ which can be used for controling output voltage disturbance as well as the input current [15], but it won't be analyzes in this paper. In terms of simulations, we used PLECS [16] with specific values $v_{IN} = 10\,\text{V}$, $L_1 = L_2 = 1\,\text{mH}$, $C_1 = 1\,\mu\text{F}$, $C_2 = 20\,\mu\text{F}$ and $R = 5\,\Omega$. Signal u has been produced using a hysteresis type of control with $\pm\Delta$, for $\Delta \in \{1\,\text{m}, 10\,\text{m}, 100\,\text{m}\}$, where m denotes $10^{-3}$ and we do not include units in $\Delta$ to simplify the notation yet it is obvious they match the corresponding variables, such as volts for voltages.

### A. Switching surface $S(\mathbf{x}) = x_1 - m_5$

Sliding surface $S(\mathbf{x}) = x_1 - m_5$ controls and limits converter's input current with coefficients $m_1 = 1$ and $m_{2-4} = 0$. The equivalent control can be computed for $x_3 > v_{IN}$ as $u_{eq} = \frac{x_3 - v_{IN}}{x_3}$, which results in the following steady-state:

$$x_1^* = m_5, \qquad x_2^* = \mp\sqrt{\frac{m_5 v_{IN}}{R}}, \quad (17)$$
$$x_3^* = v_{IN} \mp \sqrt{m_5 v_{IN} R}, \qquad x_4^* = \pm\sqrt{m_5 v_{IN} R}.$$

In order to achieve $x_3 > 0$ and $x_4 < 0$, the second solution is chosen. The first solution is negative and thus physically unacceptable.

The system's stability on the sliding surface can be verified through the introduced approach of linearization and by finding matrix $\mathbf{P}$ as in Eq. (9) to establish Lyapunov stability.

On the sliding surface $S(\mathbf{x}) = x_1 - m_5$, variable $x_1$ is constant and thus the equation for $\dot{x}_1$ can be eliminated. The stability analysis is then performed in terms of variables $\mathbf{z} = [y_2 \; y_3 \; y_4]^T$ and the remainder nonlinearity

$$\mathbf{h}^*(\mathbf{z}) = \left[ 0 \quad \frac{-v_{IN} y_2 y_3 + x_2^* y_3^2}{(y_3 + x_3^*) x_3^* C_1} \quad 0 \right]^T$$

During the switching subinterval when the switch S is on and the diode is off, the inductor current $i_{L1}$, presenting controlling variable, linearly changes, according to the linear ripple approximation:

$$\Delta i_{L1} = 2\Delta = \frac{v_{IN}}{L_1} u_{eq} T_S. \quad (18)$$

Using Eq. (25), the switching period $T_S$ can be computed and then used in the ripple calculation for other state variables as given in Eqs. (21), (22) and (23).

In Figure 2, steady state waveforms for different values of the regulator hysteresis bound $\Delta \in \{10\,\text{m}, 100\,\text{m}\}$ when $m_5 = 0.5$, are shown. Steady-state values are then $\mathbf{x}^* = [0.5\,\text{A} \; 1\,\text{A} \; 15\,\text{V} \; -5\,\text{V}]^T$.

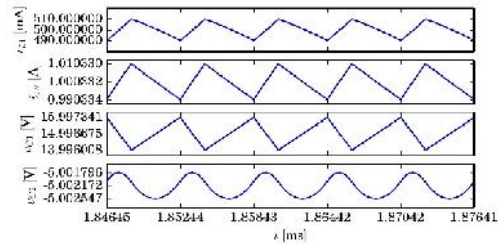

*(a) Calculated and measured values are* $T_S = 6\,\mu\text{s}$, $\Delta i_{L1} = \Delta i_{L2} = 20\,\text{mA}$, $\Delta v_{C1} = 2\,\text{V}$ *and* $\Delta v_{C2} = 750\,\mu\text{V}$.

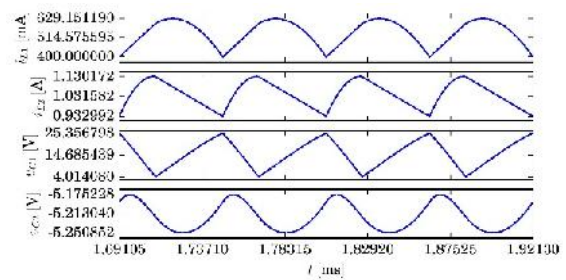

*(b) Calculated values are* $T_S = 60\,\mu\text{s}$, $\Delta i_{L1} = \Delta i_{L2} = 200\,\text{mA}$, $\Delta v_{C1} = 20\,\text{V}$ *and* $\Delta v_{C2} = 75\,\text{mV}$. *Measured is* $T_S = 57.55\,\mu\text{s}$.

*Figure 2 – Steady-state waveforms for regular hysteresis bound (a)* $\Delta = 10\,\text{m}$; *(b)* $\Delta = 100\,\text{m}$.

Furthermore, the diagrams in Fig. 2a show a high level of linear behavior and fit well the linear ripple approximation analysis. In Fig. 2b, the nonlinearity in the ripple becomes evident especially in the waveform for $i_{L1}$. It shows that the time constants caused by real parts of eigenvalues are comparable to the switching period.

The system stability is verified using the LMI approach and computing the nonlinearity bound. The nonlinear remainder $\mathbf{h}^*(\mathbf{z})$ has nonzero expression only at the location $h_3^*(y_3) = \mathbf{h}^*(y_3)$. By choosing matrix $\mathbf{H}$ to be a 3 by 3 matrix with all entries equal to zero except the one at position (2,2) being equal to 1, and using the LMI convex program as in Eq. (9) we obtain the maximized sector size. In particular, one can show that the nonlinear term satisfies $|h_3^*| \leq \left(\bar{r}/\|\hat{\mathbf{T}}_2\|\right)|y_3|$, where $\hat{\mathbf{T}}_2$ is the second column of the inverse of a transformation matrix $\mathbf{T}$ as in Eq. (14). The nonlinearity perturbation bound is computed as $\bar{r} = 9.4995 \cdot 10^3$.

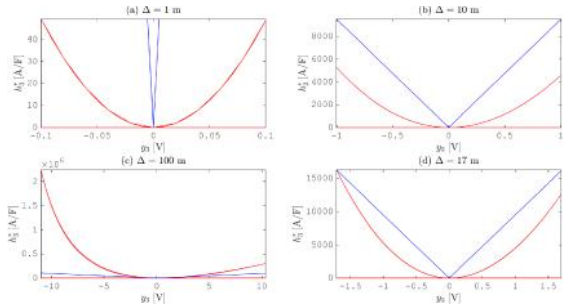

Figure 3 – The nonlinear term $h_3^*$ is plotted in red and the sector bound $\bar{r}|y_3|$ is plotted in blue for different values of  in subfigures (a)-(d).

In Figure 3, diagrams $h_3^*(y_3)$ for different values for  are shown. It can be seen that the error $h_3^*$ lies inside the perturbation limit when $\Delta \in \{1\,\text{m}, 10\,\text{m}\}$ and exceeds the sector bound when $\Delta = 100\,\text{m}$. A dominant time constant that causes nonlinearity influencing waveforms is due to eigenvalues $\}_{1,2} = \dfrac{-\sqrt{L_2} \pm \sqrt{L_2 - 4C_2 R}}{2C_2 R \sqrt{L_2}}$ of the system model when the switch S is off and the diode D is on. The case when $\Delta = 100\,\text{m}$ shows a large nonlinear behavior especially in the waveform for $i_{L1}$ when $u = 0$, as depicted in Figure 2b. It can be also observed that the left side of the diagram in Figure 3c shows steeper increasing of the error while decreasing $y_3$. In this case the error forces the system to approach the other infeasible equilibrium (the first solution in Eq. (24)).

Considering obtained sector bound  in Eq. (11), the limit for applying linear ripple approximation can be determined as crossing point of nonlinearity $h_3^*$ and the sector bound $\bar{r}|y_3|$. It can be seen that $\Delta y_3 = \Delta v_{C1} \approx 3.4\,\text{V}$ is a limit for considering linear ripple approximation in this example and from Eq. (21), the following can be derived:

$$\Delta = \frac{1}{2} \frac{v_{IN} \Delta v_{C1}}{x_2^*} \frac{C_1}{L_1}, \quad (19)$$

which provides $\Delta = 17\,\text{m}$ as a limit. In Figure 3d this is depicted as a limit ripple confirming the LMI procedure computation. Clearly the calculated error for all simulated cases shows when and if the linear ripple approximation can be used.

B. Switching surface $S(x) = m_1 x_1 + m_2 x_2 - m_5$

The sliding surface $S(x) = m_1 x_1 + m_2 x_2 - m_5$ allows hysteresis window current mode control [12] in which the average of a weighted sum of inductors' currents is kept at a desired constant value. The equivalent control produces equilibria

$$x_1^* = \frac{2m_1 m_5 R + m_2(m_2 v_{IN} \pm \sqrt{v_{IN}} E)}{2R m_1^2}, \quad x_2^* = \frac{m_5 - m_1 x_1^*}{m_2},$$

$$x_3^* = v_{IN} - \frac{m_2 v_{IN} \pm \sqrt{v_{IN}} E}{2m_1}, \quad x_4^* = \frac{m_2 v_{IN} \pm \sqrt{v_{IN}} E}{2m_1}, \quad (20)$$

for $E = \sqrt{m_2^2 v_{IN} + m_1 m_5 R}$. Of our interest is the second solution because it provides a positive value for $x_3$ and a negative value for $x_4$. This choice results in

$$u_{eq} = \frac{E - m_2 \sqrt{v_{IN}}}{\sqrt{v_{IN}}(2m_1 - m_2) + E}.$$

For the case of the switching surface $S(x) = m_1 x_1 + m_2 x_2 - m_5$ we can eliminate state variable $y_2$ and get reduced state vector $\mathbf{z} = [y_1 \; y_3 \; y_4]^T$. The nonlinear remainder $\mathbf{h}^*(\mathbf{z}) = \begin{bmatrix} 0 & h_3^* & 0 \end{bmatrix}^T$ is given by

$$h_3^* = -\frac{m_2 x_4^* y_3 - R v_{IN}(m_1 - m_2) y_1}{m_2 C_1 R v_{IN} x_3^* x_3 (m_1 L_2 + m_2 L_1)} \times$$

$$\times \left( y_3 (m_2 L_1 x_4^* + m_1 L_2 v_{IN}) - y_4 m_2 L_1 x_3^* \right). \quad (21)$$

We can compute $T_S$, which is then used to compute inductors currents' and capacitors voltages' ripple as described by Eqs. (21), (22) and (23) based on

$$2\Delta = m_1 \Delta i_{L1} + m_2 \Delta i_{L2} = \left(\frac{m_1}{L_1} + \frac{m_2}{L_2}\right) v_{IN} u_{eq} T_S. \quad (22)$$

In order to demonstrate a bound for the linear ripple approximation particular, the values for switching surface $S(x) = m_1 x_1 + m_2 x_2 - m_5$ are chosen as $m_1 = m_2 = 1$, $m_5 = 2$, which produce the following steady-state values.

$$\mathbf{x}^* = \left[(3-\sqrt{5})\,\text{A} \ \ \frac{6-2\sqrt{5}}{\sqrt{5}-1}\,\text{A} \ \ 5(\sqrt{5}+1)\,\text{V} \ \ \frac{-30+10\sqrt{5}}{\sqrt{5}-1}\,\text{V}\right]^T$$

Using the LMI convex program we obtain the sector size parameter value as $r = \tilde{r}/\|\hat{\mathbf{T}}_2\| = 7.5308 \cdot 10^3$.

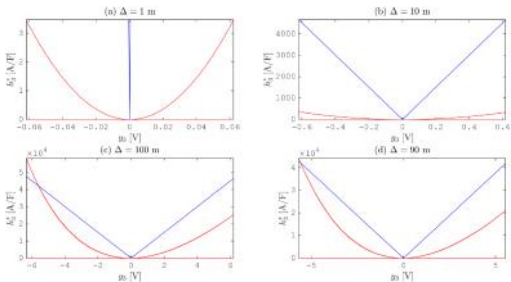

Figure 4 – The nonlinear term $h_3^*$ is plotted in red and the sector bound $r|y_3|$ is plotted in blue for different values of in subfigures (a)-(d).

The nonlinear remainder term is computed based on simulations when $\Delta \in \{1\,\text{m}, 10\,\text{m}, 100\,\text{m}\}$ and shown in Figure 4. One may notice an effect of nonlinearity when $\Delta = 100\,\text{m}$, as depicted in Figure 4c. We can also observe maximal for which the influence of the system eigenvalues is negligible comparing to the switching period and estimate $\Delta v_{C1,\max} = 11.1\,\text{V}$. By using the following expression:

$$\Delta = \frac{\Delta v_{C1} v_{IN} C_1}{2 x_2^*}\left(\frac{m_1}{L_1} + \frac{m_2}{L_2}\right) \quad (23)$$

we compute $\Delta = 90\,\text{m}$. We can confirm that this is indeed the boundary value as depicted in Figure 5d The resulting phase diagram for this sliding surface is given in Figure 5. At a start up the system's trajectory follows $x_2 = 0$ and $u = 1$, before it gets to the switching surface $m_1 x_1 + m_2 x_2 - m_5 = 0$. Then there is a small overshoot when $u = 0$ until a steady state is reached. After that the system's motion stays on $m_1 x_1 + m_2 x_2 - m_5 = 0$. When $m_5$ is large enough so that DCVM occurs, after reaching the sliding surface, then voltage $v_{C1}$ drops to zero, so the system's motion in DCVM returns to the switching surface via yellow trajectories as depicted in Figure 6 (with $u = 1$). In the case of the bidirectional realization there is no discontinuous regime, so when $m_5$ is very large the return to the switching surface occurs as depicted by green trajectories in Figure 5, that is, $u = 1$.

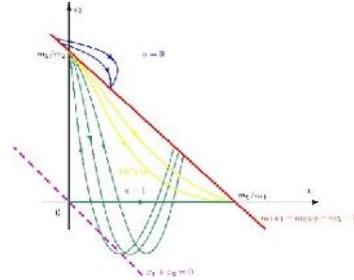

Figure 5 – Phase diagram $x_2(x_1)$ in the case when $S(x)=m_1x_1+m_2x_2-m_5$

In order to see the difference between the case of unidirectional switch S and a diode and bidirectional switches, in Figure 6 time diagrams when $S(\mathbf{x}) = x_1 + x_2 - 4$ are provided. One may also observe that DCVM reduces significantly the inductor currents' ripple and the capacitor $C_1$'s voltage. In addition, the phase diagrams shown in Fig. 7 confirm the previous analysis and agree with the phase plot from Figure 5.

## 5. CONCLUSION

In this paper we applied an equivalent control model of a sliding (or switching) surface to characterize the steady-state analysis of DC/DC converters. This is done by linearizing nonlinear sliding mode dynamics and representing it by a linear part and a sector bounded nonlinear remainder. Furthermore, using the linear matrix inequalities stability approach we estimated the size and the shape of the sector which the nonlinear remainder satisfies. The nonlinear sector bound is then used to determine the limit for applying linear ripple approximation in the converter operation analysis. This approach is demonstrated of the uk converter's example for two switching surfaces of the practical importance.

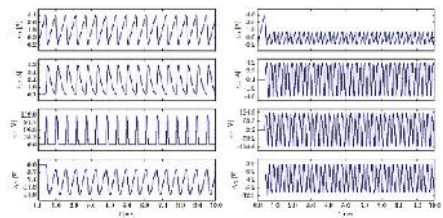

*(a)*  *(b)*

Figure 6 – State variables versus time in the case when $S(\mathbf{x}) = x_1 + x_2 - 4$ when the switches are realized as (a) an unidirectional switch and a diode; (b) bidirectional

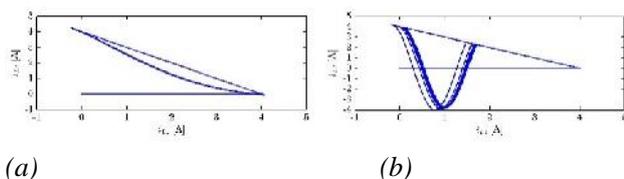

*(a)* *(b)*

*Figure 7 – Phase diagram $i_{L2}(i_{L1})$ in the case when $S(\mathbf{x}) = x_1 + x_2 - 4$ when the switches are realized*

## 6. ACKNOWLEDGEMENT

The authors would like to thank Professor Vujo Drndarevi for his generous support and valuable comments which greatly increased the quality of this paper.

REFERENCES

[1] Vadim I. U. *Survey paper variable structure systems with sliding modes*. IEEE Transactions on Automatic control. 1977 Apr;22(2):212-222.

[2] DeCarlo R, Zak SH, Matthews GP. *Variable structure control of nonlinear multivariable systems: a tutorial,* Proceedings of the IEEE. Mar; 76(3):212-32, 1988.

[3] Filippov A. F. Differential equations with discontinuous right-hand side. Matematicheskii sbornik. 93(1):99-128, 1960.

[4] El Ghaoui L, Feron E, Balakrishnan V. *Linear matrix inequalities in system and control theory,* Philadelphia: Society for industrial and applied mathematics; 1994 Jun.

[5] Boyd S, Vandenberghe L. *Convex optimization,* Cambridge university press; Mar 8 2004.

[6] Chesi G, Colaneri P, Geromel J. C, Middleton R, Shorten R, *A nonconservative LMI condition for stability of switched systems with guaranteed dwell time*, Automatic Control, IEEE Transactions on. 2012 May;57(5):1297-302.

[7] Šiljak D. D, Stipanovic D. M. *Robust stabilization of nonlinear systems: the LMI approach*, Mathematical problems in Engineering, 6(5):461-93, 2000.

[8] Huang S. P, Xu H. Q, Liu Y. F, *Sliding-mode controlled Cuk switching regulator with fast response and first-order dynamic characteristic,* In Power Electronics Specialists Conference, 1989. PESC'89 Record., 20th Annual IEEE (pp. 124-129), Jun 26 1989.

[9] Oppenheimer M, Husain I, Elbuluk M, De Abreu Garcia J. A. *Sliding mode control of the Cuk converter,* In Power Electronics Specialists Conference, 1996. PESC'96 Record, 27th Annual IEEE (Vol. 2, pp. 1519-1526), Jun 23 1996.

[10] Martinez-Salamero L, Calvente J, Giral R, Poveda A, Fossas E. *Analysis of a bidirectional coupled-inductor Cuk converter operating in sliding mode. Circuits and Systems I: Fundamental Theory and Applications,* IEEE Transactions on. 45(4):355-63, Apr 1998.

[1] Martínez-Salamero L, García G, Orellana M, Lahore C, Estibals B, Alonso C, Carrejo C. E, *Analysis and design of a sliding-mode strategy for start-up control and voltage regulation in a buck converter,* IET Power Electronics. 6(1):52-9, Jan 2013.

[2] Erickson R. W, Maksimovic D, *Fundamentals of power electronics,* Springer Science & Business Media; May 8 2007.

[3] Utkin V. *Sliding mode control of DC/DC converters,* Journal of the Franklin Institute. 31;350(8):2146-65, Oct 2013.

[4] Yakubovich V. A, *S-procedure in nonlinear control theory,* Vestnik Leningrad University, 1:62-77, 1971.

[5] Chen Z, *PI and sliding mode control of a Cuk converter*, IEEE Transactions on Power Electronics, 27(8):3695-703, Aug 2012.

[6] Alimeling J. H, Hammer W. P, *PLECS-piece-wise linear electrical circuit simulation for Simulink,* In Power Electronics and Drive Systems, 1999. PEDS'99. Proceedings of the IEEE 1999 International Conference on 1999 (Vol. 1, pp. 355-360).

**REZIME**

LMI PRSTUP U PROJEKTOVANJU SLIDING-MODE KONTROLE I ANALIZI DC-DC KONVERTORA

*Prekida ko ponašanje kola, specijalno DC-DC konvertora je analizirano u ovom radu koriš enjem ekvivalentne kontrole za modelovanje sliding-mode režima dinami kih sistema. Kao reprezentativni primer i jedan od najkompleksinijih DC-DC konvertora, izabran je uk konvertor. Pokazuje se da se ponašanje konvertora u ustaljenom stanju može posmatrati i analizirati koriš enjem uslova stabilnosti baziranih na linearnim matri nim nejedna inama sa nelinearnom perturbacijom ograni enom u sektoru. Maksimizacija nelinearnog ograni enja u sektoru daje granicu za primenu linear ripple aproksimacije u analizi rada konvertora. Štaviše, naš pristup je potvr en simulacijama za dve razli ite prekida ke površine od interesa.*

**Klju ne re i:** *DC/DC konvertor, uk konvertor, sliding-mode, linearne matri ne nejedna ine*